\documentclass[fleqn,twoside]{article}
\usepackage{espcrc2}

\usepackage{graphicx}
\usepackage[figuresright]{rotating}
\usepackage{amsmath}
\def\gsim{\raise0.3ex\hbox{$>$\kern-0.75em\raise-1.1ex\hbox{$\sim$}}}
\def\lsim{\raise0.3ex\hbox{$<$\kern-0.75em\raise-1.1ex\hbox{$\sim$}}}

\title{Lorentz violation, vacuum, cosmic rays, superbradyons and Pamir data}

\author{Luis Gonzalez-Mestres\address{LAPP, Universit\'e de Savoie, CNRS/IN2P3, B.P. 110, 74941 Annecy-le-Vieux Cedex, France}}

\begin{document}

\begin{abstract}
The possibility that Pamir data at very high energy cannot be fully explained by standard physics has recently led to the suggestion that the peculiar jet structure observed above $\sim ~ 10^{16}$ eV could be due to a suppression of effective space transverse dimensions. The new pattern considered violates Lorentz symmetry. We point out that, in models with Lorentz symmetry violation, a suppression of available transverse energy for jets while conserving longitudinal momentum can be generated by new forms of energy losses at very high energy without altering space-time structure. An illustrative example can be superbradyon emission, where in all cases the superbradyon energy would be much larger than its momentum times c (speed of light). More generally, such phenomena could be due to the interaction of the high-energy cosmic ray with new vacuum and/or particle structure below the $10^{-20}$ cm scale. Scenarios involving Lorentz symmetry violation but not superbradyons are also briefly considered.

\vspace{1pc}
\end{abstract}

\maketitle

\section{Introduction}

It has recently been suggested \cite{Landsberg1} that, if Pamir and other similar data above $\sim ~ 10^{16}$ eV \cite{mountain,stratosph} cannot be explained by standard particle physics, the peculiar jet structure observed could be due to a suppression of effective space transverse dimensions \cite{Landsberg2}. This suppression would generate elongated or elliptic shapes in high-energy jets. 

Although the situation concerning the possible need of new physics to explain such data remains unclear \cite{DeRoeck}, the subject obviously deserves further study. In particular, concerning possible similar implications and signatures of new physics in the high-energy cosmic-ray region.

The pattern suggested in \cite{Landsberg1,Landsberg2} violates Lorentz symmetry and implies the existence of an absolute rest frame similar to the vacuum rest frame (VRF) considered in our papers on Lorentz symmetry violation (LSV) and superbradyons since 1995 \cite{gonSL-LSV}. It therefore seems relevant to investigate more generally possible ways to generate elongated jets and alignement phenomena in models with LSV, and to discuss how essential is dimension folding to produce such effects.

LSV at the Planck scale \cite{Gonzalez-Mestres2008,Gonzalez-Mestres2009a} is being partially tested by ultra-high energy cosmic-ray (UHECR) experiments \cite{AugerData,HiresData}. Results from the AUGER \cite{AUG} and HiRes \cite{Hires} collaborations can possibly confirm the existence of the Greisen-Zatsepin-Kuzmin (GZK) cutoff \cite{GZK1,GZK2}, whose suppression as a consequence of LSV and deformed relativistic kinematics (DRK) was suggested in \cite{Gonzalez-Mestres97a,Gonzalez-Mestres97b}. But by now, uncertainties concerning acceleration sources prevent from reaching definite conclusions on the grounds of existing data \cite{AugerSources}. 

Furthermore, the question of a possible LSV appears correlated with that of the validity of the models used to describe the interaction of a cosmic ray with the atmosphere at ultra-high energies (UHE) and to evaluate its energy.

Even if the existence of the GZK cutoff were eventually confirmed, important LSV patterns and domains of parameters would remain allowed \cite{Gonzalez-Mestres2008,Gonzalez-Mestres2009a}. A long-term program with further UHECR experiments (Auger North \cite{AugerNorth}, Telecope Array \cite{TA}, satellite missions like JEM-EUSO \cite{JEM-EUSO}...) will therefore be required.  

It does not seem that an elongation of jets can be produced at the $10^{16}$ eV scale just by a deformation of relativistic kinematics like those considered \cite{gonSL-LSV} in models of quadratically deformed relativistic kinematics (QDRK) originating from LSV generated at the Planck scale or at some other fundamental length scale. But, as pointed out in \cite{Gonzalez-Mestres05-06}, LSV can present threshold effects at intermediate scales between the low-energy region and the Planck scale. Such thresholds would be related to a change in the basic physics involved, or to the appearance of new physical phenomena. In particular, this hypothesis would allow to reconcile stronger LSV effects at very high energy with existing low-energy bounds on LSV. The LSV pattern considered in \cite{Landsberg1,Landsberg2} is based on strong threshold
effects concerning effective space dimensions and the associated vacuum structure.

In general, we do not expect LSV phenomena occurring in very high energy cosmic ray (VHECR) interactions to produce related signatures at LHC energies. Even in the absence of a sharp threshold, the LSV parameters must decrease with the energy scale in order to be consistent with low-energy experiments \cite{Lamoreaux}. 

For similar reasons, one can question the validity of the extrapolation of standard center-of-mass interaction patterns to the collision of a VHECR with the atmosphere if Lorentz symmetry is violated. Estimates of UHECR energies and composition can therefore be uncertain in the presence of possible LSV phenomena.

It this note, we point out that missing transverse energy in cosmic-ray interaction jets above some energy scale ($\sim ~ 10^{16}$ eV ?) can be a natural consequence of the production of superluminal objects (waves, particles...) involving a small portion of energy (actually provided by the target) and a negligible fraction of momentum. Assuming a kinematics of the superbradyonic type \cite{gonSL-LSV}, a significant (longitudinal or transverse) momentum for the produced exotic objects would be forbidden by its energetic cost. 

The new mechanism, based on energy capture by vacuum from standard particle scattering, could be present up to UHECR energies and influence interactions and signatures.

Thus, in a UHECR collision with the atmosphere, the production of such superluminal objects can potentially be present and absorb a substantial part of the total target energy that would otherwise be converted into transverse energy of secondaries. Elongated jets and similar effects can possibly be generated by this mechanism. Although transverse dimensions are not suppressed, the effective transverse motion of the produced particles or waves is limited by high-momentum kinematics and energy conservation. 

\section{Lorentz symmetry violation}

Our 1997 and subsequent papers \cite{gonSL-LSV} rejected linearly deformed relativistic kinematics (LDRK), as it naturally leads to too large effects below Planck scale. A simple QDRK pattern can use the following kinematics in the VRF :
\begin{equation}
E~=~~(2\pi )^{-1}~h~c~a^{-1}~e~(k~a)
\end{equation}

\noindent where $E$ is the particle energy, $h$ the Planck constant, $c$ the speed of light, $k$ the wave vector, $a$ the fundamental length and $[e~(k~a)]^2$ a convex function of $(k~a)^2$. Expanding (1) for $k~a~\ll ~1$ , one gets \cite{Gonzalez-Mestres97a,Gonzalez-Mestres97b}:
\begin{equation}
e~(k~a) ~ \simeq ~ [(k~a)^2~-~\alpha ~(k~a)^4~+~(2\pi ~a)^2~h^{-2}~m^2~c^2]^{1/2}
\end{equation}
$p$ being the particle momentum, $\alpha $ a positive model-dependent constant and {\it m} the mass of the particle. For $p~\gg ~mc$ , one has:
\begin{equation}
E ~ \simeq ~ p~c~+~m^2~c^3~(2~p)^{-1}~-~p~c~\alpha ~(k~a)^2/2
\end{equation}

QDRK is similar to a refraction equation with a Cauchy law. It can therefore be considered as describing refraction of standard particles by the physical vacuum. It can also be seen \cite{gonSL-LSV,Gonzalez-Mestres97a} as the equivalent of the dispersion relation for phonons in condensed matter physics. 

Although QDRK is not by itself a specific theory, it describes the deformation of relativistic kinematics that can be introduced by a large set of theoretical scenarios incorporating a fundamental length scale (Planck or beyond). The deformation term ($-~\alpha ~(k~a)^4$ in [2] and $-~p~c~\alpha ~(k~a)^2/2 $ in [3] for $k~a~\ll ~1$) accounts quite generally for the expected effect below the fundamental energy and momentum scales.

LDRK models, where the deformation term would be $-~\beta ~(k~a)^3$ in [2] and $-~p~c~\beta ~k~a/2 $ in [3], will not be considered here. As stressed above, they were rejected in our original papers \cite{gonSL-LSV} for two reasons : i) they lead to too large effects for reasonable values of $\alpha $ and $a$ not too far from Planck scale ; ii) assuming that a LSV effect would be observed at the UHECR scale, a linear energy-dependence of the LSV parameter would not be enough to reproduce the successful experimental low-energy checks of Lorentz symmetry. LDRK was, however, partly incorporated in some LSV patterns exhibiting energy thresholds \cite{Gonzalez-Mestres05-06}.

The deformations considered here are basically different from strong doubly special relativity (SDSR) patterns \cite{SDSR} which assume the laws of Physics to be identical in all reference frames. The existence of the VRF is a necessary condition to get observable predictions from energy-dependent LSV for VHCR and UHECR physics. Otherwise, calculations could be performed in the center-of-mass frame where LSV and deformation effects would be much smaller. 

As specified in previous papers \cite{Gonzalez-Mestres2008,Gonzalez-Mestres2009a,Gonzalez-Mestres05-06}, we call weak doubly special relativity (WDSR) our LSV approach based on : i) a fundamental length, originating a deformation of Lorentz symmetry where standard relativity remains as a low-energy limit for conventional particles : ii) the existence of an absolute local rest frame, defined by the physical vacuum. The proposal by Anchordoqui et al. \cite{Landsberg1,Landsberg2}, and the alternative presented here, are compatible with WDSR but not with SDSR.

In what follows, calculations are systematically performed in a reference frame assumed to be close to the VRF. The VRF is usually associated to the local frame where cosmic microwave background radiation appears to be isotropic. This natural assumption implies a close connection between cosmology and vacuum dynamics.

Actually, defining the VRF may become a nontrivial task if the concept of "vacuum rest frame" itself turns out to depend on the wavelength scale. Such a possible complication will not be considered at the present phenomenological stage. For all practical purposes, we assume that : i) in the VRF, standard Lorentz symmetry remains an exact symmetry in the low-momentum limit ; ii) our laboratory system is close to the VRF.

If the VRF can be compared to the rest frame of a conventional physical object, its existence may imply that our Universe is not a unique and "total" entity but a component of a larger material medium. An important question would then be whether some form of matter and/or energy can enter or leave this Universe, just as photons can enter or leave standard condensed matter.

\section {Kinematics}

Assuming energy and momentum conservation and, for simplicity, that an incoming UHECR of mass {\it m}, energy $E$ and momentum $p$ hits a target of mass {\it M} at rest, the total available energy for secondaries produced by this collision will be $E~+~M~c^2$ . Most of this energy will be spent to fulfill the requirement of momentum conservation in the direction of the incoming cosmic ray.

With the additional simplifying hypothesis that the final state is made of two particles of mass $m'$ and longitudinal momentum $p/2$ where the longitudinal direction is taken to be that of the incoming momentum, high-energy mass terms and QDRK deformations can normally be neglected to a first approximation. 

We then get :
\begin{equation}
{p_T}^2~\simeq ~M~c~p/4 
\end{equation}   
where $p_T$ is the transverse momentum of each produced particle with respect to the direction of the incoming momentum. The transverse energy of each secondary is then $E_T~\simeq ~M~c^2/2$ , the incoming UHECR energy being entirely spent to account for the longitudinal momenta of secondaries and the available transverse energy being entirely provided by the target energy.

Therefore, the available transverse energy for secondaries can become considerably smaller in the presence of a simultaneous emission of objects with a total energy comparable to that of the target and total longitudinal momentum much smaller than the emitted energy times $c^{-1}$. 

If $\Delta E_{vac}$ is the energy lost in a process involving a momentum $\Delta p_{vac} ~<<~\Delta E_{vac} ~c^{-1}$, and assuming the same final state configuration as before, the transverse energy of each secondary will become $E_T~\simeq ~(M~c^2~-~\Delta E_{vac})/2$ , leading to :
\begin{equation}
{p_T}^2~\simeq ~(M~c~-~\Delta E_{vac} ~ c^{-1})~p/4   
\end{equation}   

\section{Superbradyons}

In standard particle physics, the only activity of vacuum occurs through the Higgs mechanism, initially inspired by condensed matter physics, where bosons condense with spontaneous symmetry breaking phenomena. But the vacuum can be a much more lively object if conventional particles are actually composite entities.

As emphasized in previous papers \cite{Gonzalez-Mestres2009b}, superluminal preons (superbradyons) with a critical speed in vacuum $c_s~\gg ~c$ can be the ultimate constituents of matter and play a fundamental role in cosmology. Superbradyons would have positive mass and energy, and be basically different from tachyons. They can obey, for instance, a new Lorentz invariance with $c_s$ as the critical speed.

The relation of the superbradyon sector, its basic physics and its fundamental quantum numbers, with the apparent properties of standard particles is expected to be far more involved than in the early preon models considered three decades ago \cite{Salam1,preons}. For this reason, and because the composite structure of conventional "elementary" particles is assumed to be generated beyond Planck scale, superbradyons are expected to couple very weakly to "ordinary" particles at energies far below the Planck energy, apart from exceptional situations and processes.

Assuming a kinematics of the Lorentz type with $c_s$ playing the role of the critical speed, the energy $E_s$ and momentum $p_s$ of a free superbradyon in the VRF would be given by : 
\begin{eqnarray}
E_s~=~c_s~(p_s^2~+~m_s^2 ~c_s^2)^{1/2} \\
p_s~=~m_s~v_s~(1 ~-~v_s^2~c_s^{-2})^{-1/2}
\end{eqnarray}
where $m_s$ is the inertial mass of the superbradyon and $v_s$ its speed. 

Superbradyons can undergo refraction in the physical vacuum, but this effect shall not be considered here. Similar to their interaction with matter, we expect superbradyon refraction in vacuum to be very small at low energy. 

The existence of the VRF in our Universe would also break locally the possible new Lorentz invariance associated to superbradyons, although this would not be an absolute breaking. This effect is neglected here. In such an approach, standard Lorentz symmetry would be just an internal property of a condensed medium (like the low-momentum Lorentz symmetry for phonons in a solid), whereas the superbradyonic Lorentz symmetry could be a real fundamental symmetry.

Obviously, one has in all cases $p_s~c~\ll~E_s$. Superbradyon emission would, basically, release energy but not momentum as compared to "ordinary" particles with the same energy. This is the natural source of kinematical constraints on the allowed superbradyon energy and momentum in physical processes involving conventional particles. It leads to $\Delta p_{vac} ~<<~\Delta E_{vac} ~c^{-1}$ for the emission of superbradyons with total energy $\Delta E_{vac}$ and total momentum $\Delta p_{vac}$, and sets a limit on the allowed energy for such a process in cosmic-ray interactions with the atmosphere.

As conservation of longitudinal momentum requires spending all the incoming cosmic-ray energy, the emission of a superbradyon in the collision of a UHECR with the atmosphere would necessarily occur at the expense of the available transverse energy for "ordinary" secondaries. Such a phenomenon may look like a suppression of effective transverse dimensions.

An elongation of UHECR jets above some energy scale can therefore be a signature of the opening of a threshold for the production of "light" superbradyons (superbradyons with comparatively low rest energy). It would possibly reflect specific phenomena related to the inner structure of the physical vacuum and of conventional particles at the distance scale associated to the incoming cosmic-ray energy ($\approx ~10^{-20}$ cm for $E~\approx ~10^{16}$ eV). The total superbradyon energy would be limited by the available transverse energy and, therefore, by the target energy (including rest energy).

Such a scenario would in particular imply that superbradyons have in any case played and important role in the early universe at temperatures above $\approx ~10^{20}$ K, and that they exist nowadays as a component of dark matter and a source of dark energy (including a possible internal vacuum evolution). This new component of superbradyonic dark matter would have rest energies much lower than those considered in \cite{Gonzalez-Mestres2009b} on the grounds of the experimental data obtained by PAMELA \cite{PAMELA} and other experiments.   

\section {Alternative scenarios}     

More generally, it may happen that the physical vacuum reacts to the occurrence of a collision involving wavelengths smaller than some critical scale, and that new physics manifests itself through this dynamical reaction to the conventional particle collision. 

Then, the energy and momentum spent by such a non-standard phenomenon do not need to be directly correlated to those of the incoming high-energy particle. Vacuum can, for instance, capture energy from the conventional particles to generate excitations related to the new physics contained in its internal structure and dynamics.

Similar unconventional phenomena could also be generated by : i) the internal particle structure (cosmic ray and/or target) excited by the very short wavelength of the incoming cosmic ray, the target providing the required energy ; ii) a combined effect of the internal vacuum and particle structure.

\subsection {Other superbradyonic processes}
\vskip 2mm

If vacuum can directly reveal its superbradyonic structure, and if superbradyonic dynamics is able to generate waves related to a zero-mass superbradyon or to another kind of vacuum excitation, the energy captured from the scattering between "ordinary" particles can possibly be dissipated in the form of concentric waves propagating isotropically from the collision point.   

Another possibility would be that the energy captured by vacuum generates an unconventional virtual item that quickly decays into superbradyonic particles and/or waves. 

Similar considerations would apply to phenomena generated by the internal structure of the conventional particles.

As they could in general travel initially at a speed larger than $c$, superbradyonic particles and waves produced by such processes are expected to spontaneously emit conventional waves and particles until they reach propagating speeds close to $c$ \cite{gonSL-LSV,Gonzalez-Mestres2008} where "Cherenkov" emission in vacuum is no longer allowed.

\subsection {Non-superbradyonic scenarios}
\vskip 2mm

Non-superbradyonic phenomena can also be considered. LSV would then be relevant only to generate the thresholds for the internal vacuum and particle structure in the VRF. 

Conventional particles would then be emitted by the excited vacuum (and/or particle) structure due to new physics unraveled below the critical distance scale, and be the decay products of the virtual objects thus produced. 

To lead to the unconventional effect considered, vacuum is assumed to basically capture energy, but not a substantial amount of momentum, from the UHECR collision with the atmosphere. Otherwise, most of the energy would have to be spent in the emission of standard high-energy particles.

\section{Conclusion and comments}

It appears that patterns violating Lorentz symmetry can potentially generate in several different ways mechanisms reproducing phenomena like jet elongation in UHECR collisions with the atmosphere above some critical energy scale. The suggestions presented in this paper are just illustrative examples.  

In the superbradyonic approach considered here, the inner vacuum structure appears as a natural candidate to generate new dynamical phenomena if it is excited below a critical wavelength scale. Then, conventional particles may just be excitations of the superbradyonic vacuum. However, these excitations may also be by themselves carriers of new (topological ?) structures able to produce relevant new physics. 

In all cases, the existence of a threshold effect and the difference in energy scale would in principle make compatible the possible new high-energy phenomena considered in this article with existing low-energy bounds on LSV \cite{Lamoreaux}. 

Patterns of interactions of the superbradyonic world with standard matter, or describing any other LSV effect below the threshold energy (including at LHC) need particular caution in order not to be in conflict with the low-energy bounds. But, for the possible phenomena considered here and in \cite{Gonzalez-Mestres2009b}, compatibility does not appear impossible when numbers are considered \cite{Gonzalez-Mestres2010}.

A superbradyon with energy in the GeV range as suggested in this paper, assuming $v_s ~ \sim ~ c$ and $c_s ~ \sim ~10^6$ $c$ (similar to the ratio between the speed of light and that of phonons), would have a mass $\sim ~ 10^{-3}$ eV $c^{-2}$, momentum $\sim ~ 10^{-3}$ eV $c^{-1}$ and kinetic energy $\sim ~ 10^{-3}$ eV.

Similarly, taking the same values for $v_s$ and $c_s$, a superbradyon with energy in the TeV range as suggested in \cite{Gonzalez-Mestres2009b} to explain the observed positron abundance \cite{PAMELA} by superbradyon annihilations or decays, would have a mass $\sim ~ 1$ eV $c^{-2}$, momentum $\sim ~ 1$ eV $c^{-1}$ and kinetic energy $\sim ~ 1$ eV. 

In both cases such superbradyons would naturally be very difficult to detect, not only because of their expected very weak interaction with conventional matter but also because of the small available energy for such interactions. 

The situation would be different if the observed positron flux were due to "Cherenkov" emission by superbradyons with $v_s$ slightly above $c$ and $c_s ~ \sim ~ 10^6$ $c$ . In this case, the superbradyon rest energy would be $\sim ~10^{24}$ eV and some spectacular decays could perhaps be observed.

The systematic test of fundamental principles by cosmic-ray physics is an essential and long-term task. It concerns not only relativity, but also \cite{Gonzalez-Mestres2009b,Gonzalez-Mestres2010} the ultimate structure of matter, quantum mechanics, energy and momentum conservation, the validity of the Hamilton - Lagrange formulation, CPT, standard cosmology... Remarkably, high-energy cosmic rays appear to offer unique and unprecedented possibilities in this domain. 

In particular, VHECR and UHECR provide a unique and direct probe of vacuum structure and internal dynamics at distance and energy scales beyond the reach of accelerator projects. Not only through the acceleration \cite{Gonzalez-Mestres2000} and interactions of the cosmic rays, but also through their propagation in vacuum over long distances. Similar considerations apply to the internal structure and dynamics of conventional particles at very short wavelength scales \cite{Gonzalez-Mestres2004}.  

The inner vacuum dynamics can be a basic ingredient of the dark energy phenomenon, driving the present evolution of the Universe by mechanisms beyond the scope of standard particle physics. Unconventional cosmic-ray phenomena may be indirect signatures of the same dynamics.

\end{document}